# Node seniority ranking

Vincenzo Fioriti[1]* and Marta Chinnici[1]

[1]ENEA, Casaccia Laboratories, via Anguillarese 301, S. Maria in Galeria, 00123, Roma, Italy

*Correspondence to: vincenzo.fioriti@enea.it

**Abstract**: Recent advances in graph theory suggest that is possible to identify the oldest nodes of a network using only the graph topology. Here we report on applications to heterogeneous real world networks. To this end, and in order to gain new insights, we propose the theoretical framework of the Estrada communicability. We apply it to two technological networks (an underground, the diffusion of a software worm in a LAN) and to a third network representing a cholera outbreak. In spite of errors introduced in the adjacency matrix of their graphs, the identification of the oldest nodes is feasible, within a small margin of error, and extremely simple. Utilizations include the search of the initial disease-spreader (patient zero problem), rumors in social networks, malware in computer networks, triggering events in blackouts, oldest urban sites recognition.

We investigate the growth over time of graphs, identifying the source nodes who started the growth on a pure topological basis. Is it feasible to classify nodes according to their age without measurements ? The common sense answer a few years ago would be, predictably, puzzling. Nevertheless, recently *(1)* it has been shown how to trace the oldest node sources of an evolving graph, using only the eigenvalues and eigenvectors of the Laplacian matrix. Such an identification task is to be regarded as a difficult one, but advantages to be gained in many fields of science are so relevant to justify the efforts of a large number of researchers. The interest of this inverse problem lies in the variety of applications in IT security, medicine, pharmacology, archaeology, finance, engineering, biology, but till a few years ago solutions were not foreseen. We show that is possible to identify oldest nodes of heterogeneous real world technological networks or of an epidemic spreading graph, namely: the underground of Paris during the period 1900-1949, the diffusion of a software worm in a computer LAN, a cholera outbreak. Moreover, we suggest a necessary condition to recognize the networks suitable of the age-analysis and a rough estimator of the algorithm performance.

Pinto *(2)* has developed a procedure to estimate the location of the epidemic source from measurements collected by sparsely placed observers by means of a maximum probability estimator. Each observer (about 20% of the nodes were monitored) measures from which neighbour and at what time has received the contagion. The collected data are used to produce the estimate, whose complexity is $O(N^3)$, $N$ number of nodes. Results of the validation test on the Kwa Zulu cholera outbreak in South Africa in 2000 show the estimation errors is below 4 hops. In this paper we consider the "patient zero" as the oldest node of the cholera outbreak graph, thus we see no difference among the three different networks and consequently we can apply the same methodology.

Zhu *(1)* instead has developed a deterministic spectral strategy based only on the topology of the network (solving *de facto* an inverse problem) at the same computational cost $O(N^3)$, applying his method to the Santa Fè co-authorship social network *(3)* and to the protein-protein interaction network.

Here we present the solution to both problems applying a similar, very simple methodology. Our main goal is to study some heterogeneous prototypical real-world networks in order to provide tools for practical applications. The graph (is the mathematical counterpart network, but the two words are almost equivalent) to be analyzed results from a growth intended as an evolution over time, depending generally on stable, non-stochastic, "smooth" transformations. When its topology is known, Zhu claims that the eigenvalue spectrum of the connectivity matrix or preferably of the Laplacian matrix is related closely to the age of nodes. The correlation between eigenvalues and age is strictly required; moreover, if no evolutionary process was developed in the past, the method is not applicable. For graphs following the preferential attachment rule ("rich get richer") the correlation is clear, because the probability for a node to acquire new links is proportional to its degree, therefore a strong correlation between the node degree and its life-time is sure, but real world networks are much more complicated *(4 -6)*. For a given eigenvalue, the lifetime of the associated eigenvector is the average age of all nodes contained in the vector, weighted by the respective components of the eigenvector. The first step is to build the Laplacian matrix **L** = **D** – **A**, where **D** is the degree diagonal matrix and **A** the adjacency matrix ($a_{ij}$ = 1 if the link *i-j* exists, 0 otherwise ). The second step is the standardization of each eigenvector components: $v_i = | v_i / max (v_i) |$, with $i = 1, 2, ... N$. The third step is the seniority ranking. Nodes with standardized component values larger than a threshold are clustered in a certain age subset and related to the associated eigenvalues, thus the largest eigenvalues is associated to the oldest node and so on. This method, tested on the Santa Fè Institute co-authorship of scientific papers social network *(3)*, is able to classify the age of nodes *(1)* completely. For example, the first three larger eigenvalues of the Laplacian, related to the nodes corresponding to the eigenvectors selected by the thresholding procedure, indicate the three oldest nodes of the network of Fig. 1: $\lambda_{76}>\lambda_{75}>\lambda_{74} \leftrightarrow$ **40, 7, 67** where **40** is the oldest node, and the $\lambda_N>\lambda_{N-1} > \ldots >\lambda_1$ is the descending eigenvalue spectrum. The Zhu' procedure is due to the observation that the eigenvector size in networks, such as the protein-protein interactions, do not seems to increase, while the corresponding (according to the threshold procedure) eigenvalue does. No suggestions about the characteristics of evolving networks suitable to be age-analyzed or how to choose the threshold's value are given. Now we note that has been discovered independently *(7)*, in many social networks, how in large non random graphs changes over time result in a continuous variation of the adjacency matrix eigenvalues, while the eigenvectors stay (relatively) constants, therefore, the correlation between the node ages and the largest eigenvalues comes as a direct consequence. Although authors of *(7)* apparently were not aware of the relation eigenspectrum-age, they sketch a demonstration for a necessary condition on the eigenvectors, that we consider a sound approach to explain the age – eigenvalue correlation, as follows. Starting from the standard eigenvalue decomposition of a graph: $\mathbf{A}(t_i) = \mathbf{V}(t_i)\, \Lambda(t_i)\, \mathbf{V}(t_i)$ $i = 0, 1, 2, \ldots , N$ where **A** is the adjacency matrix, **V** the eigenvectors matrix, **V'** its transpose, $\Lambda$ the eigenvalues matrix, at time $t_i$ .If the eigenvectors remain constants, we can write: $\mathbf{A}(t_{i+1}) \approx$ $\mathbf{V}(t_i)\, \Lambda(t_{i+1})\, \mathbf{V}(t_i)'$ where $\Lambda(t_{i+1}) = \Delta(\mathbf{A}, \mathbf{V}, \Lambda) + \Lambda(t_i)$. Since **V** has orthogonal columns we can compute the best fit of $\Delta$ in a least-squares sense

$$\Delta'(\mathbf{A}, \mathbf{V}, \Lambda) \approx \mathbf{V}(t_i)(\mathbf{A}(t_{i+1}) - \mathbf{A}(t_i))\mathbf{V}(t_i)'$$

and since the calculation requires $\Delta'$ to be diagonal, any deviation from this condition indicates a deviation from the type of graph evolution over time required, deteriorating the age evaluation. Actually, the diagonality condition may be relaxed to a diagonal dominance. A note of caution: for random graphs, such as Erdos-Renyi graphs, eigenvectors increase faster than eigenvalues *(7)*, hence the age analysis is unfeasible. Now we introduce our alternative procedure. We have seen $\mathbf{A}(t_{i+1}) \approx \mathbf{V}(t_i)\, \Lambda(t_{i+1})\, \mathbf{V}(t_i)'$ then $tr(e^{\mathbf{A}(ti+1)}) = tr(\mathbf{V}(t_i) e^{\Lambda(ti+1)}\, \mathbf{V}(t_i)') = \Sigma_j\, e^{\lambda j}$, for $\lambda_j=\lambda_j(t_i+1)$. If eigenvectors stay almost constants, most of the variation of the trace from time $t_i$ to $t_i$+1 depends from the eigenvalues; in particular, each node $i$ contributes with the quantity $SC_i = \Sigma_j\, (v^i_j)^2 e^{\lambda j}$, where $\lambda_j=\lambda_j(t_i+1)$ and $v^i_j$ denote the $i$-th component of the eigenvector $\boldsymbol{v}_j$. The $SC_i$ parameter *(8)* called sub-graph centrality, is closely related to the communicability index ECI defined *(11)* as:

$$\mathbf{ECI} = e^{\mathbf{A}}$$

where $i = j$ determines the diagonal entries of **ECI** matrix that are the $SC_i$ values, while for $i \neq j$ we have communicability between node $i$ and node $j$. Now, the larger the $ECI_{ii}$ value, the older the corresponding node $i$. Thus, sorting the diagonal entries of the **ECI** matrix is possible to recover at the same time the node number and its age-rank. Note that $SC_i$ may be regarded as a self-communicability index *(11)*, so we expect similar capabilities for both parameters. A probabilistic interpretation may be also given: $SC_i$ is proportional to the probability of a random walker passing close to node $i$. The Estrada indexes communicability and sub-graph centrality take into account not only the immediate effects of the closest nodes, but also the long-range effects transmitted through the participation of a node in all sub-graphs *(8, 9, 11)* travelling along all the paths available. This explains why ECI and SC are able to retain the information about the oldest nodes through many sub-graphs, during the time evolution. Since many important results have been established about the spectrum of the adjacency matrix *(12 - 15)*, it would be useful to use the adjacency matrix instead of the Laplacian, without losing insights about the node ages. For example, the spectrum of the adjacency matrix eigenvalues has been used in the last years to reveal the most vulnerable nodes to the epidemic spreading of viruses and malware (*13, 16)*. A unified procedure based on the eigenspectrum would be elegant, theoretically sound and could be set in the larger framework of the graph entropy, the quantum mechanics, the non linear oscillators *(10, 11)*. Then what are the advantages and drawbacks of the ECI procedure compared to the Zhu algorithm ? From an algorithmic point of view, the Estrada communicability is simpler: does not need thresholds and the information on the nodes is easily recovered as the diagonal entries of the **ECI** matrix. On the other hand, the Zhu algorithm is certainly more accurate and usually faster, unless particular parallelization techniques are used *(18)*. Thus when precision for all nodes is needed, we suggest to resort to the Zhu methods, otherwise the ECI may be considered, according to circumstances.

In order to validate the ECI procedure against benchmarks before the actual use, we have selected the social network of the Santa Fè Institute scientific co-authorship collaborations *(1, 3)*, some artificial Barabasi-Albert graphs *(6)* and the cholera Kwa Zulu outbreak *(2)*. The Santa Fé

Institute collaboration example shows that our ECI procedure recovers exactly the first two nodes (**40**, **7**) out of the three (**40, 7, 67**) oldest, as follows (Fig. 2):

node (ordered according to seniority): **40** **7** 24

ECI value: 125.19 123.78 55.75

but makes an error when tries to classify node **67** as the third oldest node. The Zhu' algorithm *(1)* in this example is able to calculate exactly the seniority for all nodes, taking full advantage from the Laplacian matrix and therefore is more accurate, however we point out that our interest is limited on the very first oldest nodes.

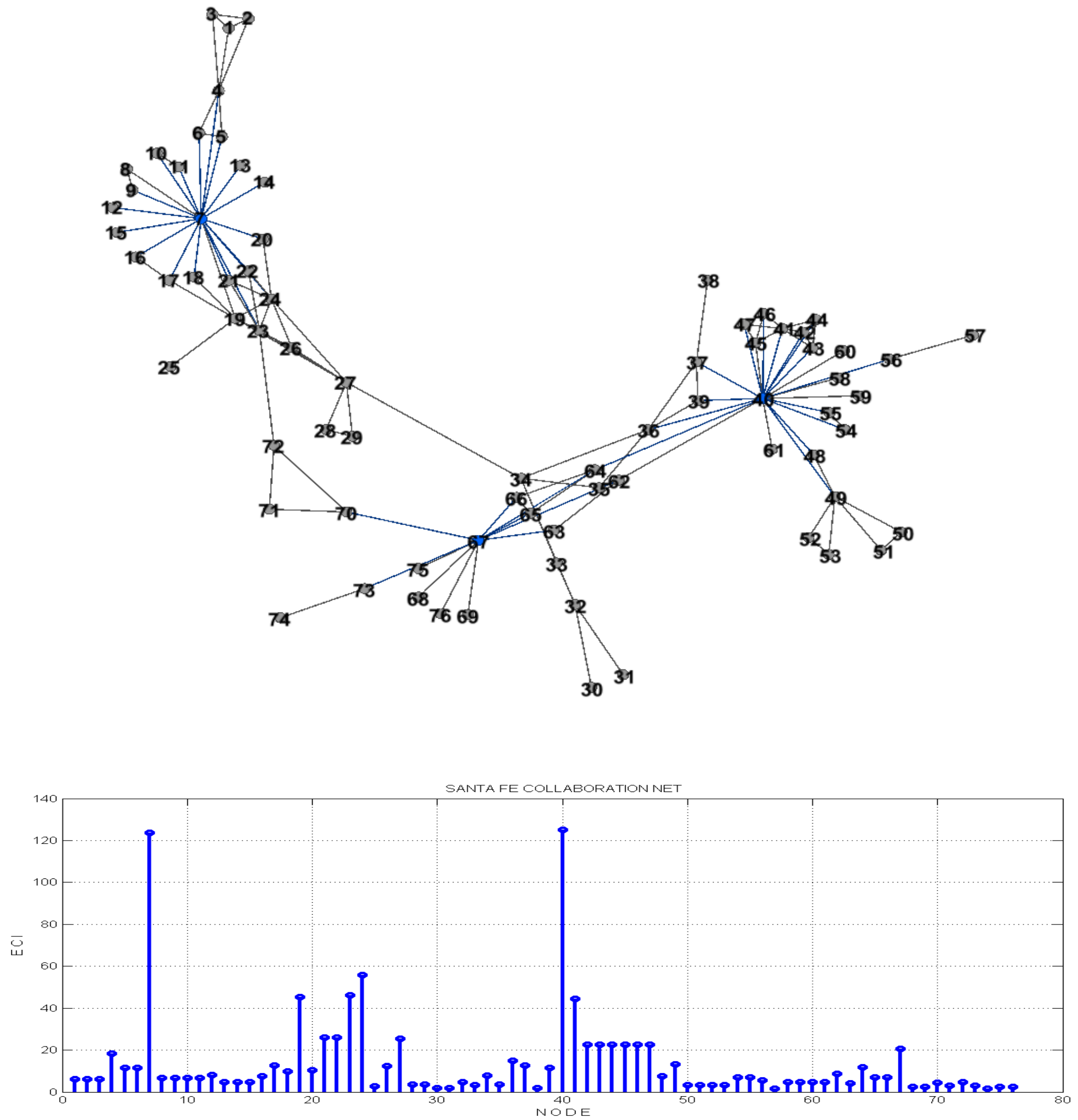

Fig. 1. (**A**) Santa Fè Institute co–authorship collaboration network. Nodes represent authors of scientific papers related to the Santa Fe Institute. Nodes **40, 7**, **67** (blue, at the centre of the major hubs) are the first, second and third oldest node, respectively.(**B**) ECI classifies the oldest nodes of the Santa Fè Institute co-authorship collaboration network. On the abscissa are the node numbers, on the ordinate the ECI values; node **7** and node **40** have both an ECI value about 120, that is the maximum value, therefore

they are the first two oldest node. ECI classifies correctly the first two (**40, 7**) out of three (**40, 7**, **67**) oldest nodes, but fails with node **67**, mistaken with node 24. Note in the red dotted circle a group of coetaneous nodes.

Another benchmark is the Barabasi-Albert graph (Fig. 3) for 1000, 2500, 10000 nodes. Locating the sources in this kind of graph is easy because of the preferential attachment rule sets a strong correlation with the degree *(1).* The ECI procedure, in fact, finds the four sources (nodes **1, 2, 3, 4**) within the first six positions of the calculated ranking (**2, 4, 1**, 21, 17, **3**), adding two false positive nodes, 21 and 17; better results with a BA 2500 nodes graph: the four sources (nodes **1, 2, 3, 4**) are within the first five positions of the ranking (**4**, 6, **3, 2, 1**) adding as a false positive only node 6, and finally for a BA graph of 10000 nodes we obtain all the sources (**2**, **3**, **1**, **4**) with no errors.

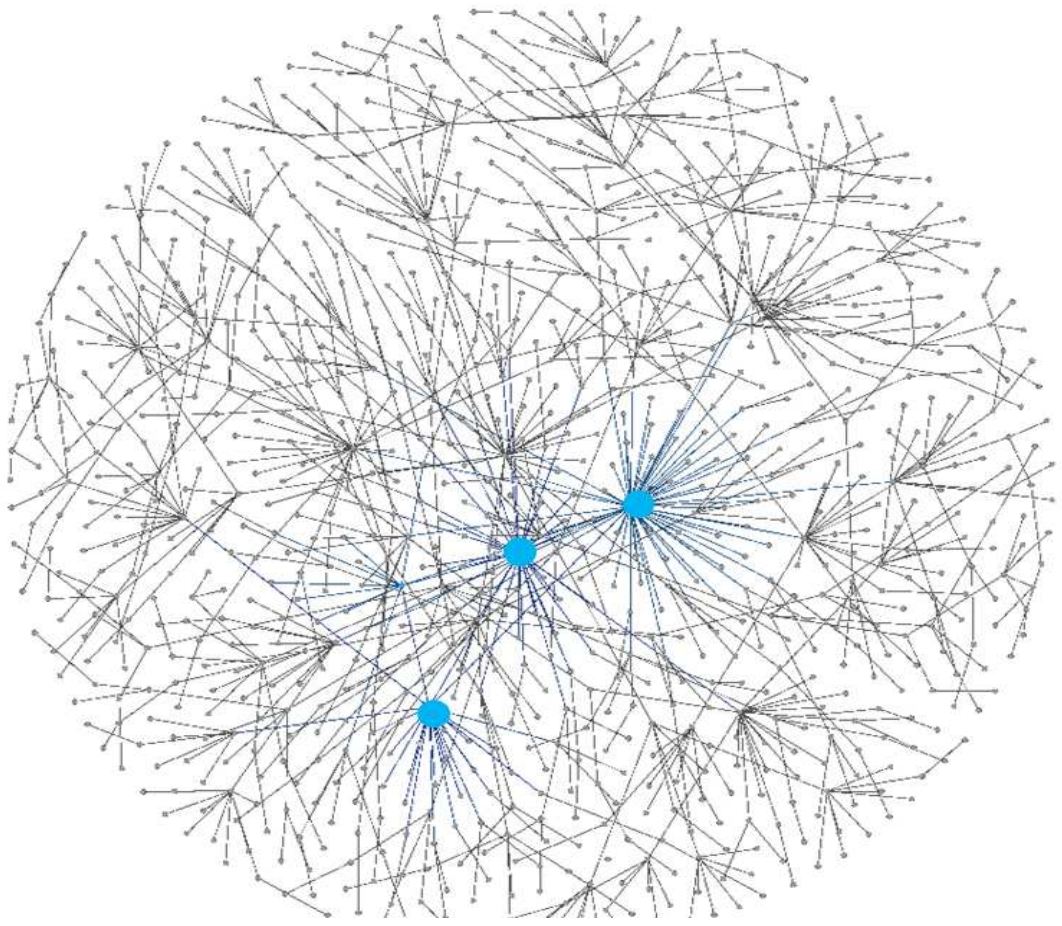

Fig. 2. A Barabasi-Albert (1000 nodes) graph. Blue large dots are the sources **1, 2, 3, 4**.

Then we conclude that the ECI method is less accurate with respect to the Zhu's algorithm, nevertheless provides a good performance for the very oldest nodes.

An important question to weigh up is whether the node seniority algorithms are robust to errors occurring in the adjacency matrix, i.e. nodes/links missing or wrongly added. In fact, very often when investigating the real world phenomena, one is compelled to face incomplete information about the topology of the network and the reliability of the algorithm becomes a major issue. In a few words, we consider that in some practical situations one would prefer to have a less precise but robust algorithm, therefore it may be convenient to stop the analysis at the very first oldest nodes, let's say 10% of the nodes. Having validate the ECI procedure now we apply it to two *(17, 16)* technological networks (the underground of Paris and a computer network) and to the graph obtained from the cholera outbreak of Kwa Zulu *(2)*, because from the point of view of the theoretical approach described before there is no difference among the three networks.

The Paris underground *(17)* during the period 1900-1949 is showed in Fig. 4. It can be clearly seen a sort of ring surrounding the downtown city with the first 1900 − 1910 underground stations. The graph has been produced considering only the most important stations and the final destinations as actual nodes (Fig. 5). The task, to identify the five oldest node (period 1900 − 1906) located inside the ring, is made more difficult by some young nodes and links added inside the ring during the period 1939 – 1949, see Fig. 4.

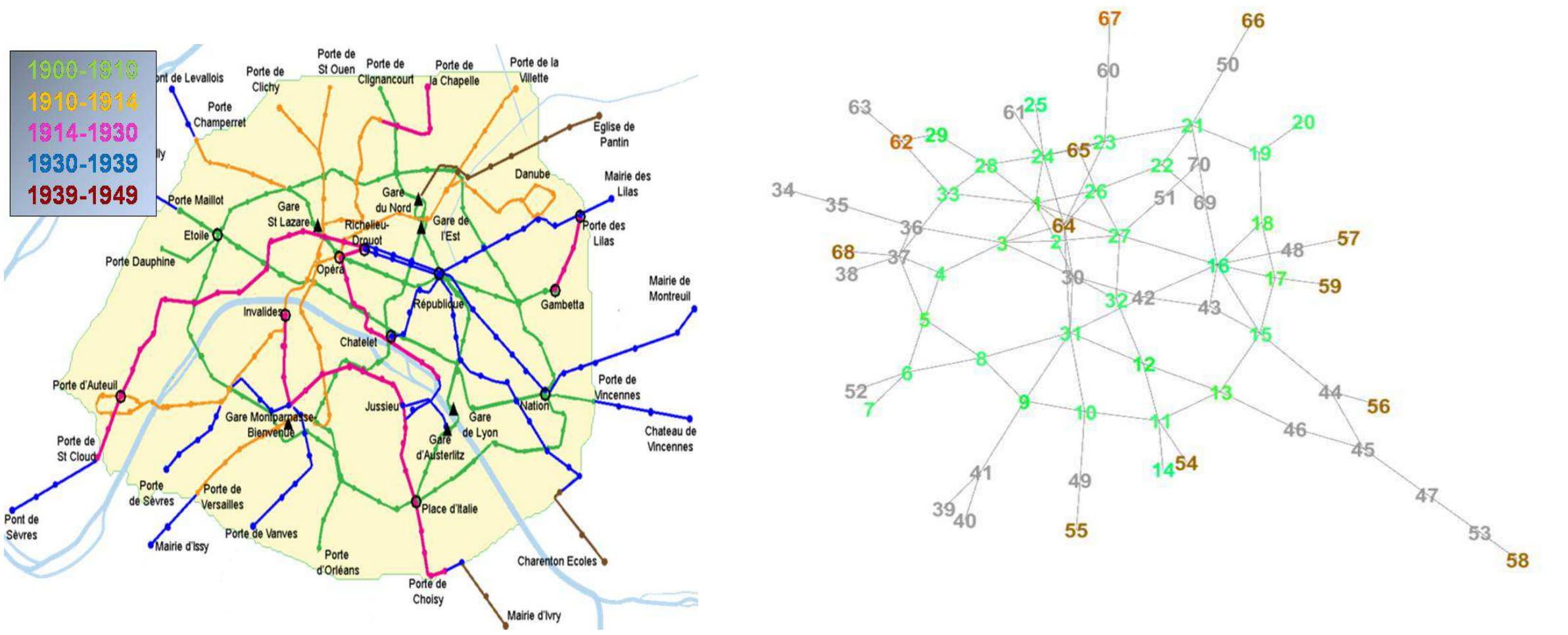

Fig. 3. The network of the underground of Paris. (**A**) The green line is the older and inside the green ring there are the very first stations dating back to the years 1900-1906 (black circles). (**B**) The graph representing the network of the main underground stations. The green nodes belong to the 1900-1910 line. The most old nodes (**1**, **3**, **16**, **27**, **31**) in the ring have been identified correctly (circle). Brown nodes are the 1939-1949 stations: most of them are the end of lines, but 64 and 65 are inside the ring (black arrows). The total number of nodes is 70.

The five oldest nodes are captured by ECI within the first eight (bold): **1**, 64, 24, **3**, 30, **31**, **16**, **27**. After 1910 many new lines were connected to the 1900-1910 nodes, disturbing the original topology; the newer structures were superimposed to the primitive ones causing a strong noise for the algorithm, thus this test is rather challenging, as always when the network is a technological one. Moreover, less important stations have been discarded, probably producing an extra amount of noise. To verify further the effect of deleting nodes we eliminate the last seven 63-70 (four of them 64, 65, 69, 70 are inside the ring causing noise); ECI captures the five actual oldest nodes **1, 16, 31, 27, 3** exactly as the first five.

But what happen if some of the oldest nodes disappear from the graph ? Eliminating the first seven nodes 1-7 (remember **1** and **3** are among the oldest five), ECI captures **16, 31**, 64 out of the remaining **16, 31, 27,** thus only node 64 is mistaken as false positive. This last result is particularly important because demonstrates that a *negative growth* (i. e. some of the oldest nodes disappearing at the end of the evolution of the graph) does not damage the age identification unduly.

In Figure 6 we have the cholera outbreak case: here we consider the actual source, node **87,** as the oldest because it developed the epidemic. The case of the cholera outbreak of Pinto shows that his probabilistic algorithm is accurate within less than four hops *(2)*. Considering only 3 hops, an exhaustive search would mean to visit fifteen nodes (82, 83, 84, 85, 86, 88, 89, 90, 92, 99, 100, 101, 102, 103, 205). On the other hand, the ECI algorithm it captures node **87** as the 14th in the calculated seniority ranking (68, 20, 140,18, 121, 124, 67, 196, 28, 27, 133, 197, 24, **87**). Therefore performances are similar, however the probabilistic algorithm of Pinto must monitor about the 20% of the nodes, instead ECI needs only the topology, that is obviously a pivotal advantage. The *ex post* (meaning the stop criterion is known) search for "patient zero" is reduced to about the 7% of the nodes. In this case, applying the ECI to the Laplacian matrix instead of the adjacency matrix would improve considerably the performance; in fact node **87**, the epidemic source, would be ranked as the 7th oldest. Moreover, we have deleted some less relevant nodes to test the robustness of the algorithm.

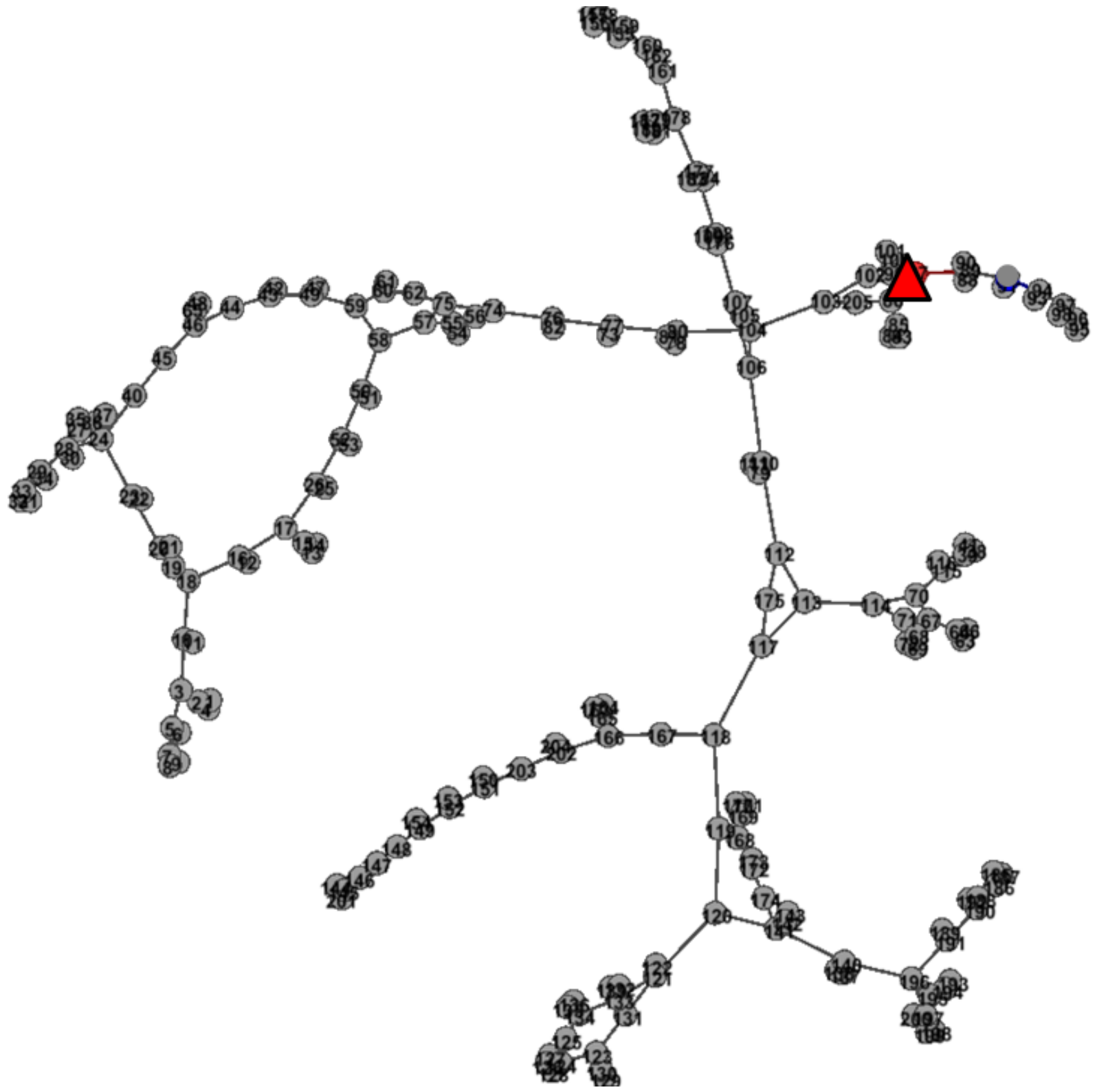

Fig. 4. Kwa Zulu cholera outbreak 2000 graph. The red triangle is the actual epidemic source, the green dot is the Pinto's estimation, within 4 hops error margin (at best, this means to visit 15 nodes). The ECI algorithm captures the correct node **87** as the 14th oldest node and does not use any measurement.The total number of nodes is 205.

The last application is the diffusion of a software worm on a computer LAN (local area network) of 759 nodes (Fig. 5). The graph suffers from the absence of a number of links, due to the inherent difficulty of the data collection *(16)*. Considering the first three ECI nodes 359, 492, 214, we find that within an error margin of 5 hops or less from 359, 214, (excluding node 492 that is completely mistaken) the actual sources **1**, **2**, **3** are all reachable. Fig. 8 shows how the sources are reachable from node 359 within 5 hops and within 4 hops from node 214. On the other hand, ECI captures two sources **3, 2** (out of **1, 2, 3)** respectively as the 25th and 27th (359, 492, 214, ... **3**, 59, **2**). The other option to find the sources is the exhaustive visit of the 27 nodes 359, 492, 214, ... **3**, 59, **2,** that is very short (3.6% of the total number of nodes).

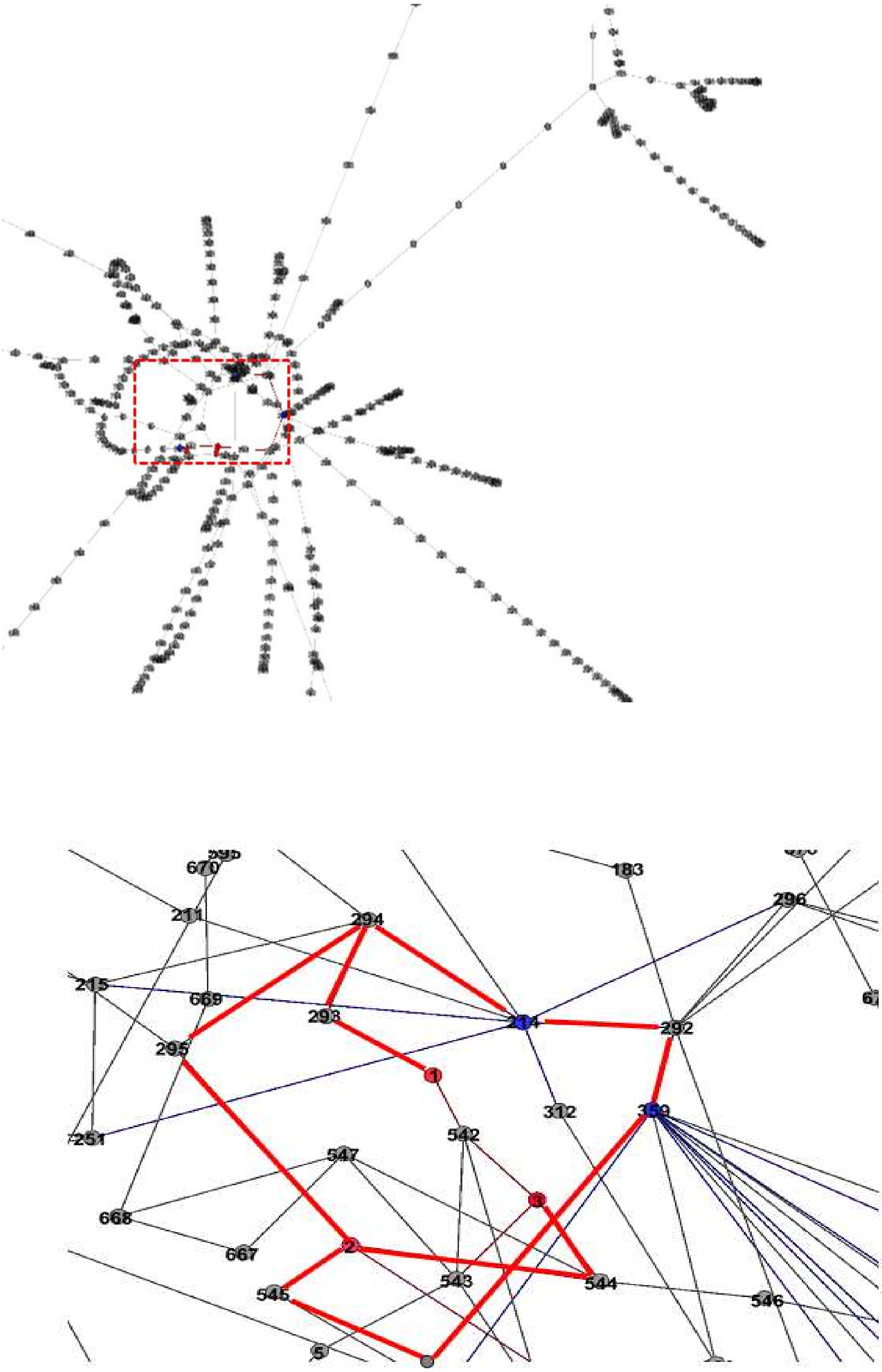

Fig. 5. (**A**) The graph of the computer LAN infected by a software worm. Infection begins from nodes **1, 2, 3** (red) inside the dotted rectangle. The network is incomplete as many links are missing, but the algorithm proves to be robust. (**B**), inset from (**A**). Note the paths (marked red links) connecting the first ranked ECI nodes 359, 214 with the red nodes (actual sources **1, 2, 3**). All sources are within 5 hops from node 359 and within 4 hops from node 214. In this case applying the ECI to the Laplacian makes the performance worse.

Of course it would be useful to know the error margin in order to stop the visit as soon as possible. We do not calculate the error margin, but provide a rough estimate of the algorithm performance, we propose a well known global index *(8, 9)*: EIN = $^1/_N\ \Sigma_i\ e^{\lambda i}$ , averaged over the number of nodes. If the eigenvalues are the algebraic counterpart of the graph geodesic properties, they should be able to provide an indication about the algorithm effectiveness, since EIN is known to be a measure of the *global* graph connectivity that influences the communicability. So a high value of EIN could be correlated to a good performance of the algorithm (cfr. Table 1). Consequently, if the EIN parameter is large with respect to a BA network of the same size, it is probably possible to stop the visit to the first 10% of the total number of nodes. Of course, this is only a preliminary analysis on a limited data set preventing statistically significant claims. To validate a true statistical correlation between EIN and the outcome of the algorithm it would be necessary an extensive analysis on a very large number of real world networks of different kinds, supported by the standard tests.

| **Graph** | **N.ro nodes** | **ESTRADA INDEX (AVERAGED)** | **Evaluation of the performance** |
|---|---|---|---|
| LAN | 759 | 2.58 | 8 |
| ERDOS-RENYI 200 | 200 | 2.59 | 9 |
| KWA ZULU | 205 | 2.68 | 6 |
| BA 110 | 110 | 2.78 | 7 |
| UNDERGROUND | 70 | 5.55 | 5 |
| SANTA FE COLL. | 76 | 13.76 | 4 |
| BA 1000 | 1000 | 7.87 | 3 |
| BA 2500 | 2500 | 19.5 | 2 |
| BA 10000 | 10000 | 697 | 1 |

Table 1. The networks are ranked according to EIN. There are discrepancies, nevertheless the EIN follows the ECI performance. The performance ranking in the last column (the best evaluated performance is numbered 1) is somewhat arbitrary.

It was already known that the eigenspectrum describes effectively the deep characteristics of graphs, but is amazing to unveil its capabilities to identify the age of nodes on a simple topological basis in real world networks.

**Acknowledgements**:
V. F. thanks A. Fioriti, S. Ginevri, C. Iafrate, M. Ruscitti, N. Sigismondi, for useful discussions.